\documentclass[twocolumn,showpacs,preprintnumbers,amsmath,amssymb]{revtex4}
\newcommand{\be}{\begin{equation}}
\newcommand{\ee}{\end{equation}}

\newcommand{\nnbb}{\nonumber\\}

\newcommand{\bwt}{\begin{widetext}}
\newcommand{\ewt}{\end{widetext}}
\newcommand{\norm}{2\pi\hbar}
\newcommand{\norma}{2\pi}
\usepackage{epsf}
\usepackage{graphics}

\begin{document}

\title{Transport equation for the photon Wigner operator in
non-commutative QED}

\author{F. T. Brandt$^a$, Ashok Das$^b$ and J. Frenkel$^a$}
\affiliation{$^a$ Instituto de F\'{\i}sica, Universidade de S\~ao
Paulo, S\~ao Paulo, SP 05315-970, BRAZIL}
\affiliation{$^b$Department of Physics and Astronomy, University
of Rochester, Rochester, NY 14627-0171, USA}

\bigskip

\bigskip

\date{\today}

\bigskip

\begin{abstract}
We derive an exact quantum equation of motion for the photon 
Wigner operator in non-commutative QED, which is gauge covariant. 
In the classical approximation, this reduces to a simple transport
equation which describes the hard thermal effects in this theory.
As an example of the effectiveness of this method we show that,
to leading order, this equation generates in a direct way the
Green amplitudes calculated perturbatively in quantum field theory
at high temperature.
\end{abstract}


\maketitle

\section{Introduction}

Classical transport equations for point particles, in the absence of
collisions (scattering), have been well studied in the past. In more
recent years, they have also been of considerable use in the study of
hot QCD plasma \cite{kapusta:book89lebellac:book96das:book97}.
It is known, in particular, that in QCD the
leading behavior of the $n$-point gluon functions at temperatures
$T\gg p$, where $p$ represents a typical external momentum, is
proportional to $T^{2}$ and these leading contributions to the
amplitudes are all gauge 
independent \cite{Braaten:1990it,Kobes:1990dc,frenkel:1991ts}.
In order to extract the
leading order contributions to the amplitude leading to gauge
invariant results for physical quantities, it is necessary to
perform a resummation of hard thermal loops, which are defined by
\begin{equation}
p\ll k\sim T,
\end{equation}
where $k$ denotes a characteristic internal loop momentum. Such a
procedure, however, is quite
technical and the classical transport equation has provided a much
simpler method for deriving the gauge
invariant amplitudes as well as the effective action which
incorporates all  the effects of
the hard thermal loops
\cite{Heinz:1985yq,Jackiw:1993zr,Nair:1993rx,Blaizot:1994be,kelly:1994ig,Litim:sh}. 
In such an approach, one pictures the
constituents of the plasma as classical particles carrying color
charge and interacting in a self-consistent manner. 

The basic idea behind the transport equation is to determine the evolution
equation for the distribution function. There are basically two
equivalent ways of doing this. In the first
approach one 
pictures the  thermal particles, moving in an internal loop, as
classical particles in equilibrium in the hot plasma whose
dynamics are governed by classical point particle equations. The
transport equation can, of course, be derived in a straightforward
manner once we know the dynamical equations for such a particle in the
background of a gauge field. For example, let us assume that the
equations of motion for a particle in the presence of a background
field is given by
\begin{subequations}\label{force}
\begin{equation}\label{forcea}
m \frac{dx^{\sigma}}{d\tau} = k^{\sigma}
\end{equation}
\begin{equation}\label{forceb}
m \frac{dk^{\sigma}}{d\tau} =  e X^{\sigma}
\end{equation}
\end{subequations}
where $\tau$ denotes the proper time of the particle and $X^{\sigma}$
represents  the force it feels in the presence of a background
gauge field. (In addition, for a particle carrying color charge in
QCD, we have to supplement the above equations with the evolution
equation for the color charge.) The explicit form of $X^{\sigma}$ will, of
course, be different depending on the type of interaction. In
general, however, the form of $X^{\sigma}$ must be such that
$k^{2}=k^{\mu}k_{\mu}$ is a constant and that the time evolution
for $k^{\mu}$ transforms covariantly under a gauge
transformation. Given these equations, the classical transport
equation for the distribution function follows in a straightforward
manner.

In the second approach, also known as the Wigner function approach,
one starts with the Wigner distribution function for the quantum field
theory of interest interacting with a background gauge field. The
evolution equation for the Wigner function is, then, determined
directly from the equations satisfied by the quantum
fields \cite{Heinz:1985yq,Elze:1989un}.
In the case of self-interacting non-Abelian gauge fields,
there are two particular issues that need special care. First, the
Wigner function has to be defined in a gauge covariant manner and
second, for a self-interacting gauge field, one has to use a
self-consistent mean field approach to identify the background gauge
field. We will discuss this method more in detail later. But, the
advantage of the second method is that here we do not have to know the
dynamical equations governing the constituent objects moving through a
hot plasma. In both these approaches, once the transport equation for
the distribution function is known, a current is defined in terms of
it. This current, which is a functional of the background gauge
fields, generates the leading hard thermal loop amplitudes for the
theory of interest. 

\bigskip

More recently, following from developments in string theory, there
has been an increased interest in quantum field theories defined on
a non-commutative manifold satisfying
\cite{Seiberg:1999vs,Fischler:2000fv,Arcioni:1999hw,Landsteiner:2000bw,
Szabo:2001kg,Douglas:2001ba,Chu:2001fe,VanRaamsdonk:2001jd,Bonora:2000ga,Brandt:2002aa,Brandt:2002rw} 
\begin{equation}
[x^{\mu},x^{\nu}] = i \theta^{\mu\nu}\label{xmoyal}
\end{equation}
where $\theta^{\mu\nu}$ is assumed to be a constant anti-symmetric
tensor with the dimensions of length squared.
The behavior of hot plasmas in such a non-commutative
gauge theory and a corresponding transport equation for such theories
are interesting questions to study. The difficulty, on the other hand,
lies in the fact that non-commutative field theories inherently
describe particles with an extended 
structure \cite{Minwalla:1999px,Sheikh-Jabbari:1999vm,Bigatti:1999iz}
for which classical
dynamical equations are not well understood. In an earlier 
paper \cite{Brandt:2002qk},
we had tried to use the explicit forms of the leading hard thermal loop
amplitudes in non-commutative QED as well as the relation between hard
thermal loops and classical transport equations to derive a force law
for such constituents and we had proposed a transport equation for
non-commutative QED. Such a ``phenomenological'' equation
already exhibits interesting features associated with such
theories. For example, we have argued that, while the charged
particles, in such theories, have the expected dipole structure, the
charge neutral photon field exhibits a quadrupole nature in the hard
thermal loop approximation and that to describe the correct hard
thermal loop amplitudes in the leading order, the classical transport
equation must necessarily involve ``collision'' terms arising from the
extended nature of the constituent particles. 

The drawback of our earlier study lies in the fact that the
transport equation did not have a theoretical derivation. As we have
alluded to above, this is connected with the difficulty that we do not
understand well the classical dynamical equations for extended
particles in a given background. As a result, in this paper, we make an
attempt to try to derive theoretically a classical transport
equation based on the Wigner function approach, which involves only
the properties of the non-commutative QED. In Sec {\bf II}, we
describe briefly the properties of non-commutative QED and define
the Wigner distribution function for such a theory. We study various
properties of this function and derive the transport equation
associated with it. We also define the current associated with such a
system. In Sec {\bf III}, we evaluate in a direct way the
leading order amplitudes following from the current, 
which agree completely with the hard thermal loops calculated in
perturbative quantum field theory.
We also compare various features arising in this approach with
those found earlier and present a brief conclusion in Sec {\bf IV}. In
appendix {\bf A}, we give some essential technical details of the
derivation  of the transport equation. In appendix {\bf B}, we compile
some relations that are useful in understanding some other aspects of
our calculations.

\section{Wigner function approach in non-commutative QED}

Non-commutative QED is described by the Lagrangian density of the form
\be\label{Lym}
{\cal L} = - \frac{1}{4} F_{\mu\nu}\star F^{\mu\nu}
\ee
where
\begin{equation}
F_{\mu\nu} =
\partial_{\mu}A_{\nu} - \partial_{\nu}A_{\mu} - \frac{ie}{\hbar c}
\left[A_{\mu},A_{\nu}\right]_{\rm MB}
\end{equation}
where $c$ represents the speed of light. Here the Moyal bracket is
defined to be
\begin{equation}
\left[A,B\right]_{\rm MB} = A\star B - B\star A
\end{equation}
where the Gr\"{o}newold-Moyal star product has the form
\begin{equation}
A(x)\star B(x) = \left.e^{\frac{i}{2}
\theta^{\mu\nu}\partial_{\mu}^{(\zeta)}\partial_{\nu}^{(\eta)}}
A(x+\zeta)B(x+\eta)\right|_{\zeta=0=\eta}
\end{equation}
We can, of course, add to this Lagrangian density charged matter
fields, but since this theory is self-interacting, much like Yang-Mills
theory, we will continue with this. Furthermore, since the photon is
charge neutral, this case is of more interest, since we normally have
transport equations for charged particles. We simply note here that
the theory in (\ref{Lym}) is invariant under the non-commutative
$U(1)$ gauge transformation
\begin{equation}
A_{\mu} (x) \rightarrow \Omega^{-1}(x)\star A_{\mu} \star \Omega (x) +
\frac{i\hbar c}{e} \Omega^{-1} (x)\star \partial_{\mu}\Omega
(x)\label{gaugetfn}
\end{equation}

The Wigner distribution function is an operator density function in
the mixed 
phase space. For a quantum scalar field theory in four space-time
dimensions, for example, it is conventionally defined as
\begin{equation}
W(x,k) = \int \frac{d^{4}y}{(\norm)^{4}}\,e^{-\frac{i}{\hbar}
y\cdot k}\,\phi^{\dagger} (x+\frac{y}{2}) \phi
(x-\frac{y}{2})\label{wigner} 
\end{equation}
For later convenience, let us define
\begin{equation}
x_{\pm} = x \pm \frac{y}{2}\label{xpm}
\end{equation}
The definition in (\ref{wigner}) can be easily generalized to the
non-commutative scalar field theory, by simply introducing star
products in the product of fields (and treating $y$ as a parameter not
subject to the star product). In defining the Wigner function for the
non-commutative photon fields, we have to worry about the gauge
covariance properties. Since the non-commutative photon fields are
self-interacting much like the usual gluon fields in QCD, we define an
analogous gauge covariant Wigner function as
\begin{equation}
W_{\mu\nu} (x,k) = \int \frac{d^{4}y}{(\norm)^{4}}\,e^{-\frac{i}{\hbar}
y\cdot k} \,G^{(+)}_{\mu\lambda} (x)\star G^{\lambda\,(-)}_{\;\nu}
(x)\label{wignerfunction}
\end{equation}
where we have defined
\begin{equation}
G^{(\pm)}_{\mu\nu}(x) = U (x, x_{\pm})\star F_{\mu\nu} (x_{\pm})\star
U(x_{\pm}, x)\label{gmunu}
\end{equation}
and the $U$'s represent the link operators in the non-commutative
theory defined along a straight path, namely
\begin{equation}
U(x, x_{\pm}) = {\rm P}
({\rm e}_{\star}^{\large{\mp\frac{ie}{\hbar c}
\int_{0}^{1} dt\,\frac{y}{2} \cdot A (x \pm (1-t)\frac{y}{2})}})\label{link}
\end{equation}
Here ``P'' stands for path ordering from left to right and it is
straightforward to check that under a gauge transformation,
(\ref{gaugetfn}), the link operators in (\ref{link}) transform
covariantly as
\begin{equation}
U(x, x_{\pm}) \rightarrow \Omega^{-1} (x)\star U (x, x_{\pm})\star
\Omega (x_{\pm})\label{linktfn}
\end{equation}
much like in the usual Yang-Mills theory. As a result, the Wigner
function, (\ref{wignerfunction}), transforms covariantly in the
adjoint representation under a gauge transformation,
\begin{equation}
W_{\mu\nu} (x,k) \rightarrow \Omega^{-1} (x)\star W_{\mu\nu}
(x,k)\star \Omega (x)
\end{equation}

We note here that, for a link operator defined along a straight path,
we can also write
\begin{equation}
G^{(\pm)}_{\mu\nu} (x) = \left({\rm e}_{\star}^{\pm \frac{y}{2}\cdot
D}\,F_{\mu\nu} (x)\right)\label{compact}
\end{equation}
where the covariant derivative is defined to be in the adjoint
representation. Furthermore, from the definition of the Wigner function in
(\ref{wignerfunction}), it is easy to check that
\begin{equation}
W^{\dagger}_{\mu\nu} (x,k) = W_{\nu\mu} (x,k)
\end{equation}
In non-commutative QED, charge conjugation is defined as the simultaneous
transformation \cite{Sheikh-Jabbari:2000vi}
\begin{equation}
A_{\mu} (x)\rightarrow - A_{\mu} (x),\qquad \theta^{\mu\nu}\rightarrow
- \theta^{\mu\nu}\label{chargeconjugation}
\end{equation}
Under charge conjugation, it is seen, using various identities for
star products, that
\begin{equation}
W_{\mu\nu} (x,k)\rightarrow W^{c}_{\mu\nu} (x,k) = W_{\nu\mu}
(x,-k)\label{chargeconjugation1}
\end{equation}
Note from the definition in (\ref{wignerfunction}) that in the limit
$\theta\rightarrow 0$, the Wigner function for usual QED is charge
conjugation invariant, while it transforms non-trivially in the
non-commutative case. 

The derivation of the transport equation for the Wigner function, from
its definition in (\ref{wignerfunction}), follows in a straightforward
manner. We discuss the details of such a derivation in appendix 
{\bf A}.  Here, we simply note that the crucial relations that play an
important role in this derivation are given by
\bwt
\begin{eqnarray}
D^{(x)}_{\mu} U(x, x_{\pm}) & = & \partial^{(x)}_{\mu} U(x, x_{\pm}) -
\frac{ie}{\hbar
c} \left(A_{\mu} (x)\star U(x, x_{\pm}) - U(x, x_{\pm})\star A_{\mu}
(x_{\pm})\right) \nonumber\\  & = & 
\mp \frac{i}{2} y^{\nu} \left(\int_{0}^{1}
dt\,\left(e_{\star}^{\pm \frac{ty}{2}\cdot D} F_{\mu\nu}
(x)\right)\right)\star U (x, x_{\pm}) \nonumber\\ 
D^{(x)}_{\mu} U(x_{\pm}, x) & = & \partial^{(x)}_{\mu} U(x_{\pm}, x) -
\frac{ie}{\hbar c}\left(A_{\mu} (x_{\pm})\star U(x_{\pm}, x) -
U(x_{\pm}, x)\star A_{\mu} (x)\right) \nonumber\\ & = & 
\pm \frac{i}{2} y^{\nu} U (x_{\pm}, x)\star \left(\int_{0}^{1}
dt\,\left(e_{\star}^{\pm
\frac{ty}{2}\cdot D} F_{\mu\nu} (x)\right)\right)\label{relations}
\end{eqnarray}
With the relations in (\ref{relations}), it is straightforward to show
that the Wigner function for the non-commutative photon satisfies
\begin{eqnarray}\label{O22}
k \cdot D\, W_{\mu\nu} (x,k) & = & \frac{e}{2c} \frac{\partial}{\partial
k_{\sigma}} k^{\rho} 
\left[\left(\int_{0}^{1}
dt\,\left({\rm e}_{\star}^{\frac{i\hbar\,t}{2}\partial_k\cdot D} F_{\rho\sigma}
(x)\right)\right)\star W_{\mu\nu} (x,k) 
+ W_{\mu\nu} (x,k)\star
\left(\int_{0}^{1} dt\,
\left({\rm e}_{\star}^{-\frac{i\hbar\,t}{2}\partial_k\cdot D} 
F_{\rho\sigma} (x)\right)\right)
\right.\nonumber\\ & - &  \left.
\int \frac{d^{4}y}{(\norm)^{4}}\,e^{-\frac{i}{\hbar} y\cdot k}\,
G^{(+)}_{\mu\lambda} (x)\star\left(\int_{0}^{1}
dt\left(\left({\rm e}_{\star}^{\frac{ty}{2}\cdot D} F_{\rho\sigma}
(x)\right) + \left({\rm e}_{\star}^{-\frac{ty}{2}\cdot D} F_{\rho\sigma}
(x)\right)\right)\right)\star G^{\lambda\,(-)}_{\;\nu}
(x)\right]\nonumber\\
 & + &  k^{\rho} \int \frac{d^{4}y}{(\norm)^{4}}\,e^{-\frac{i}{\hbar}
y\cdot k}\left[U(x,x_{+})\star (D^{(x_{+})}_{\rho}F_{\mu\lambda})
(x_{+})\star U(x_{+}, x)\star G^{\lambda\,(-)}_{\;\nu} (x)
\right.\nonumber\\ &  &  \left.
\qquad\qquad\qquad\qquad\quad  +
G^{(+)}_{\mu\lambda} (x)\star U(x, x_{-})\star
(D^{(x_{-})}_{\rho}F^{\lambda}_{\,\nu}) (x_{-})\star U(x_{-},x)\right]
\label{equation}
\end{eqnarray}
\ewt
Here $\partial_k$ represents the derivative with respect to $k$ and
we have used the compact notation introduced in (\ref{compact})
to write the equation in a simpler form.

Equation (\ref{equation}) represents the full transport equation for the
photon Wigner function in non-commutative QED. We note that by using
various identities (which we discuss in the appendix A) the second term
in (\ref{equation}) can be written as
\begin{widetext}
\begin{eqnarray}
 &  & -\frac{ie\hbar}{4c} \frac{\partial}{\partial k_{\sigma}} \int
 \frac{d^{4}y}{(\norm)^{4}}\,e^{-\frac{i}{\hbar} y\cdot
 k}\,\int_{0}^{1} dt\,t\left\{\left[\left({\rm e}_{\star}^{\frac{ty}{2}\cdot
 D} F^{\rho}{\,\sigma} (x)\right), U(x, x_{+})\star
 (D_{\rho}F_{\mu\lambda} (x_{+}))\star U(x_{+}, x)\right]_{{\rm MB}}\star
 G^{\lambda\,(-)}_{\;\nu} (x)\right.\nonumber\\
 &  & \quad + U(x, x_{+})\star (D_{\rho}F_{\mu\lambda} (x_{+}))\star
 U(x_{+}, x)\star \left[\left(e^{-\frac{ty}{2}\cdot D}
 F^{\rho}_{\,\sigma}\right), G^{\lambda\,(-)}_{\;\nu}
 (x)\right]_{{\rm MB}}\nonumber\\
 &  & \quad + \left[\left({\rm e}_{\star}^{\frac{ty}{2}\cdot D}
 F^{\rho}_{\,\sigma} (x)\right), G^{(+)}_{\mu\lambda}
 (x)\right]_{{\rm MB}}\star U(x, x_{-})\star (D_{\rho}F^{\lambda}_{\,\nu}
 (x_{-}))\star U(x_{-}, x)\nonumber\\
 &  & \quad \left.+ G^{(+)}_{\mu\lambda} (x)\star
 \left[\left({\rm e}_{\star}^{-\frac{ty}{2}\cdot D} F^{\rho}_{\,\sigma}
 (x)\right), U(x, x_{-})\star (D_{\rho}F^{\lambda}_{\,\nu}
 (x_{-}))\star U(x_{-},x)\right]_{{\rm MB}}\right\}\nonumber\\
 &  & + \frac{e}{c} \int
 \frac{d^{4}y}{(\norm)^{4}}\,e^{-\frac{i}{\hbar} y\cdot k}\left\{U(x,
 x_{+})\star \left[F_{\mu\sigma} (x_{+}), F^{\sigma}_{\,\lambda}
 (x_{+})\right]_{{\rm MB}} \star U(x_{+}, x)\star G^{\lambda\,(-)}_{\;\nu}
 (x)\right.\nonumber\\
 &  & \quad \left. - G^{(+)}_{\mu\lambda} (x)\star U(x, x_{-})\star
 \left[F^{\lambda\sigma} (x_{-}), F_{\sigma\nu} (x_{-})\right]_{{\rm MB}}
 \star U(x_{-}, x)\right\}\label{higherorder}
\end{eqnarray}
\end{widetext}
In the hard thermal loop approximation, as we have mentioned earlier,
it is assumed that 
the gradients in the system are small compared with  $k/\hbar$.
Moreover, for the semi-classical picture to hold, one also assumes that
the ensemble average of the covariant derivative,
$\langle D\, W\rangle$,
may be considered as being sufficiently small compared with
$\langle k\, W \rangle /\hbar$. 
In this approximation, it may be verified that the 
ensemble average of the term (\ref{higherorder}), 
namely, of the second term in (\ref{equation}) is
small compared to that of the first term. 
Under the above conditions, the exponentials 
${\exp_\star}{(\pm\frac{t}{2}y\cdot D)}\sim 
{\exp_\star}{(\pm\frac{i\hbar\,t}{2}\partial_k\cdot D)}$ 
which appear in the first term of (\ref{O22}) 
may also be approximated by $1$.
Therefore, if we are only interested in
the leading order behavior in the hard thermal loop approximation, we
can neglect the second term in (\ref{equation}) and the classical
transport equation for the Wigner function takes the simple form
\bwt
\begin{eqnarray}
k \cdot D W_{\mu\nu} (x,k) & = & \frac{e}{2c} \frac{\partial}{\partial
k_{\sigma}} k^{\rho} 
\Bigl[F_{\rho\sigma}(x)\star W_{\mu\nu}(x,k) + 
      W_{\mu\nu}(x,k)\star F_{\rho\sigma}(x)
\Bigr.\nonumber\\ &  & \quad \Bigl. 
- 2\,\int \frac{d^{4}y}{(\norm)^{4}}\,e^{-\frac{i}{\hbar} y\cdot k} 
G^{(+)}_{\mu\lambda}(x)\star F_{\rho\sigma}(x)
\star G^{\lambda\,(-)}_{\;\nu}(x)\Bigr]\label{leadingequation}
\end{eqnarray}
\ewt

Since non-commutative QED is a self-interacting theory, very much like
the conventional QCD, one has to consistently separate the gauge field
into a background part and a quantum part. Normally, this is done by
assuming a mean field decomposition of the form
\begin{equation}
A_{\mu} (x) = \bar{A}_{\mu} (x) + a_{\mu} (x)\label{decomposition}
\end{equation}
where it is assumed that, in this mean field approximation,
\begin{equation}
\langle A_{\mu} (x)\rangle = \bar{A}_{\mu} (x),\qquad \langle a_{\mu}
(x)\rangle = 0
\end{equation}
In making such a decomposition, it is assumed, as is the case in the
usual  background field method, that under a gauge transformation,
(\ref{gaugetfn}), 
\begin{subequations}\label{gaugetfn1}
\begin{eqnarray}\label{gaugetfn1a}
\bar{A}_{\mu} (x) & \rightarrow & \Omega^{-1} (x)\star \bar{A}_{\mu}
(x)\star \Omega (x)\nonumber\\
& + & \frac{i\hbar c}{e}\,\Omega^{-1} (x)\star \partial_{\mu}\Omega (x),
\end{eqnarray}
\begin{equation}\label{gaugetfn1b}
a_{\mu} (x)\rightarrow \Omega^{-1}
(x)\star a_{\mu} (x)\star \Omega (x)
\end{equation}
\end{subequations}
This is very important in understanding the gauge transformation
properties of the quantities resulting from the transport
equation. We note that, under such a decomposition,
\begin{eqnarray}
F_{\mu\nu} (x) & = & \bar{F}_{\mu\nu} (x) + \bar{D}_{\mu} a_{\nu} (x) -
\bar{D}_{\nu} a_{\mu} (x) \nonumber\\
& - & \frac{ie}{\hbar c} \left[a_{\mu} (x),
a_{\nu} (x)\right]_{{\rm MB}}\label{decomposition'}
\end{eqnarray}
where
\begin{subequations}
\begin{equation}
\bar{D}_{\mu} a_{\nu} = \partial_{\mu}a_{\nu} - \frac{ie}{\hbar c}
\left[\bar{A}_{\mu}, a_{\nu}\right]_{{\rm MB}},
\end{equation}
\begin{equation} 
\bar{F}_{\mu\nu} =
\partial_{\mu}\bar{A}_{\nu} - \partial_{\nu} \bar{A}_{\mu} -
\frac{ie}{\hbar c} \left[\bar{A}_{\mu},\bar{A}_{\nu}\right]_{{\rm MB}}
\end{equation}
\end{subequations}
It is clear that every term in (\ref{decomposition'}) transforms
covariantly under the gauge transformation (\ref{gaugetfn1}).

Let us also define
\begin{equation}
{\cal G}_{\mu\nu} (x,k) = \langle W_{\mu\nu} (x,k) \rangle -
\bar{W}_{\mu\nu} (x,k)\label{decomposition1}
\end{equation}
where $\bar{W}_{\mu\nu} (x,k)$ represents the Wigner function
associated with the background field and it is important to recognize
that
\begin{equation}
\langle W_{\mu\nu} (x,k) \rangle \neq \bar{W}_{\mu\nu} (x,k)
\end{equation}

We note that $\bar{W}_{\mu\nu} (x,k)$ satisfies the same equation as
in (\ref{leadingequation}) to leading order with all background
fields. If we now assume, as is conventionally done in QCD, that all
correlations factors vanish unless they involve $G^{(\pm)}_{\mu\nu} (x)$ or
$W_{\mu\nu} (x,k)$, then we can write a transport equation for 
${\cal G}_{\mu\nu} (x,k)$ to leading order as
\bwt
\begin{eqnarray}\label{31}
k\cdot \bar{D} {\cal G}_{\mu\nu} (x,k) & = & \frac{e}{2c}
\frac{\partial}{\partial k_{\sigma}} k^{\rho} 
\Bigl[\bar{F}_{\rho\sigma}(x)\star{\cal G}_{\mu\nu}(x,k) + 
         {\cal G}_{\mu\nu}(x,k)\star \bar{F}_{\rho\sigma}(x)
\Bigr.\nonumber\\  &  & \Bigr. 
-2\int\frac{d^{4}y}{(\norm)^{4}}\,e^{-\frac{i}{\hbar}y\cdot k}
\left(
\left\langle G^{(+)}_{\mu\lambda}(x)\star\bar{F}_{\rho\sigma}(x)
\star G^{\lambda\,(-)}_{\;\;\nu}(x)\right\rangle
- \left(G^{(\pm)} (x)\rightarrow \bar{G}^{(\pm)}(x)\right)\right)
\Bigr].
\end{eqnarray}
It also follows from this, that
\begin{equation}
{\cal F} (x,k) = \frac{1}{k^{2}}\,\eta^{\mu\nu} {\cal
G}_{\mu\nu}(x,k)\label{fdefinition}
\end{equation}
would satisfy the equation
\begin{eqnarray}
k\cdot \bar{D} {\cal F} (x,k) & = & 
\frac{e}{2c} \frac{\partial}{\partial k_{\sigma}} k^{\rho} 
\Bigl[\bar{F}_{\rho\sigma}(x)\star{\cal F} (x,k) 
     +{\cal F} (x,k)\star \bar{F}_{\rho\sigma}(x)
\Bigr.\nonumber\\ &  & \quad \Bigr. 
-2
\int\frac{d^{4}y}{(\norm)^{4}}\,e^{-\frac{i}{\hbar}y\cdot k}
\frac{1}{k^{2}}\left\{\left\langle G^{(+)}_{\mu\lambda}(x)
\star\bar{F}_{\rho\sigma}(x) \star
G^{\lambda\mu\,(-)}(x)\right\rangle
- \left(G^{(\pm)} (x)\rightarrow \bar{G}^{(\pm)}
(x)\right)\right\}\Bigr]\label{fequation}
\end{eqnarray}
\ewt
This equation is manifestly covariant under the gauge transformation
(\ref{gaugetfn1}) and can be solved order by order in powers of ``$e$'' to
give the ensemble average of the Wigner function.

Given the ensemble average of the Wigner function, we can now define a
covariantly conserved current as follows. First, we note that the
current is a four vector which is  odd under charge conjugation
(\ref{chargeconjugation}). Recalling the properties of the Wigner
function under charge conjugation, (\ref{chargeconjugation1}), let us define
\begin{eqnarray}
J_{\mu} (x) & = & \frac{e}{2}\int d^{4}k\,\theta (k^{0})\,\frac{k_{\mu}}{k^{2}}
\eta^{\lambda\rho} \left({\cal G}_{\lambda\rho} (x,k) - {\cal
G}_{\rho\lambda} (x,-k)\right)\nonumber\\
& = & \frac{e}{2} \int d^{4}k\,\theta
(k^{0})\,k_{\mu}\left({\cal F} (x,k) - {\cal F}
(x,-k)\right)\label{current}
\end{eqnarray}
where it is assumed that we sum over the helicity states. 
We note that this current is manifestly odd under charge
conjugation. It transforms covariantly under the gauge transformation
(\ref{gaugetfn1}) and furthermore, using (\ref{fequation}), is easily
seen to be covariantly conserved,
\begin{equation}
\bar{D}_{\mu}J^{\mu} (x) = \partial_{\mu}J^{\mu} - \frac{ie}{\hbar c}
\left[\bar{A}_{\mu}, J^{\mu}\right]_{{\rm MB}} = 0\label{conservation}
\end{equation}
This can, therefore, be defined as the current associated with our
system and can be determined to any order in powers of ``$e$'' once
${\cal F} (x,k)$ is determined. This is a functional of $\bar{A}_{\mu}
(x)$ and would generate $n$-point amplitudes through functional
derivation. These can then be compared with the leading order
calculations from perturbation theory in the hard thermal loop
approximation.

\section{Leading order amplitudes}

Given the equation for ${\cal F} (x,k)$, (\ref{fequation}), we can now
determine it order by order in ``$e$''. In this section, we will
calculate this quantity explicitly to second order which would give us
photon  amplitudes up to the three point function for which we have
explicit results from the calculations in perturbation theory. 
For simplicity of notation, we shall use in the following natural
units $c=\hbar=1$.

To zeroth order, ${\cal F}^{(0)}$ can be calculated from its definition
in (\ref{fdefinition}) using Eqs.. (\ref{wignerfunction}) and
(\ref{decomposition1}), 
\bwt
\begin{equation}
{\cal F}^{(0)} (x,k) = \int
\frac{d^{4}y}{(\norma)^{4}}\,e^{-i y\cdot
k}\,\frac{2}{k^{2}}\,\langle (\partial_{\mu}a_{\lambda}
(x_{+})-\partial_{\lambda} a_{\mu} (x_{+}))\star
\partial^{\lambda}a^{\mu} (x_{-})\rangle\label{f0}
\end{equation}
\ewt
The thermal ensemble average in (\ref{f0}) can be calculated in a
standard manner using the field decomposition for the quantum field
$a_{\mu}$. The only important thing to remember is that we want
manifest gauge covariance preserved in the calculation. This suggests,
as in the background field method, that the proper gauge condition on
the quantum fields should maintain this invariance and, in particular,
a transverse gauge has to be generalized to the form
\begin{equation}
D_{\mu} a^{\mu} (x) = 0\label{gauge condition}
\end{equation}
This would suggest that the polarization tensors, in such a case, need
not be transverse to the momentum four vector and, in fact, can have a
longitudinal component depending on the background field. Fortunately,
up to the leading three point amplitudes, this modification does not
give rise to any contribution and, for all practical purposes of our
calculations in this paper, we can take the polarization to be
transverse to the momentum four vector. In higher order calculations
or in the sub-leading terms, however, one has to include such
contributions carefully.

With this observation, we note that Eq. (\ref{f0}) can be evaluated in
a straightforward manner and the thermal contribution has the form
\bwt
\begin{eqnarray}
{\cal F}^{(0)} (x,k) & = & \int
\frac{d^{4}y}{(\norma)^{4}}\,e^{-i y\cdot k}\,
\int \frac{d^{4}\tilde{k}}{(2\pi)^{3}}\,\theta (\tilde{k}^{0})\delta
(\tilde{k}^{2})n_{B} 
(\tilde{k}^{0}) \left(-2\sum_{s} \epsilon_{\lambda}
(\tilde{k},s)\epsilon^{\lambda} 
(\tilde{k},s)\right) \left(e^{i\tilde{k}\cdot x_{+}}\star
e^{-i\tilde{k}\cdot x_{-}} + e^{-i\tilde{k}\cdot x_{+}}\star
e^{i\tilde{k}\cdot x_{-}}\right)\nonumber\\
 & = & \frac{4}{(2\pi)^{3}} \delta (k^{2}) n_{B}
(|k^{0}|)\label{f0final},
\end{eqnarray}
where $n_{B}(k_0)\equiv(\exp{(k_0/T)}-1)^{-1}$.

We can now substitute ${\cal F}^{(0)}$ into (\ref{fequation}) to
determine ${\cal F}^{(1)}$. 
However, the calculation is not quite as
iterative in the Wigner function approach, as it normally is in the
other way of doing. We still have to evaluate some terms on the right
hand side of (\ref{fequation}) that do not involve ${\cal F}$ directly. 
In trying to evaluate such terms, we note that, to the
leading order, we can set the link operators to unity in these
terms. This is easily seen from the fact that if we expand the link
operators to any order in ``$e$'', they will involve powers of $y$
which can be thought of as $\frac{\partial}{\partial k}$ acting on the
integral. Each power of $y$, therefore, gives a contribution that is
more and more sub-leading (since $k$ is large) and, consequently,
if we are interested only in the leading contributions, we can
approximately set the link operators to unity in the second group of
terms on the right hand side of (\ref{fequation}). With this, it
follows that to lowest order in ``$e$'', these terms give a
contribution of the form
\begin{equation}
-e \frac{\partial}{\partial k_{\sigma}} k^{\rho} \int
 \frac{d^{4}y}{(\norma)^{4}}\,e^{-i y\cdot
 k}\frac{2}{k^{2}} \langle (\partial_{\mu}a_{\lambda} (x_{+}) -
 \partial_{\lambda}a_{\mu} (x_{+}))\star \bar f_{\rho\sigma} (x)\star
 \partial^{\lambda}a^{\mu} (x_{-})\rangle\label{f1}
\end{equation}
Here, we have identified the Abelian part of the field strength tensor
as
\begin{equation}
\bar f_{\rho\sigma} (x) = 
\partial_{\rho}\bar   A_{\sigma} (x) - 
\partial_{\sigma}\bar A_{\rho}   (x)\label{abelianpart}
\end{equation}
The thermal ensemble average in (\ref{f1}) can be evaluated in a
straightforward manner and the temperature dependent part has the form
\begin{eqnarray}
 & {} & -e\frac{\partial}{\partial k_{\sigma}} k^{\rho} \int
 \frac{d^{4}y}{(\norma)^{4}}\,e^{-i y\cdot k}\int
 \frac{d^{4}\tilde{k}}{(2\pi)^{3}}\,\theta(\tilde{k}^{0}) \delta
 (\tilde{k}^{2}) n_{B}
 (\tilde{k}^{0})\left(-2\sum_{s}\epsilon_{\lambda}
(\tilde{k},s)\epsilon^{\lambda}
 (\tilde{k},s)\right)\nonumber\\
 &  & \quad \times \left(e^{i\tilde{k}\cdot x_{+}}\star \bar f_{\rho\sigma}
 (x)\star e^{-i\tilde{k}\cdot x_{-}} + e^{-i\tilde{k}\cdot x_{+}}\star
 \bar f_{\rho\sigma} (x)\star e^{i\tilde{k}\cdot x_{-}}\right)\nonumber\\
 & = & -\frac{4e}{(2\pi)^{3}} \frac{\partial}{\partial k_{\sigma}}
 k^{\rho}
 \left(\delta (k^{2}) n_{B} (|k^{0}|) 
\bar f_{\rho\sigma}(x+\theta\, k)\right)\label{f1'}
\end{eqnarray}
\ewt
Here, we have used the star product identity
\begin{equation}
e^{ik\cdot x}\star f(x)\star e^{-ik\cdot x} = f(x+\theta k)\label{identity}
\end{equation}
with the identification
\begin{equation}
(\theta k)^{\mu} = \theta^{\mu\nu} k_{\nu}\label{identity1}
\end{equation}

With the determination of the second term on the right hand side of
(\ref{fequation}) to the lowest order, we can now determine 
${\cal F}^{(1)}$ from (\ref{fequation}) as
\bwt
\begin{equation}
k\cdot \partial {\cal F}^{(1)} (x,k) - ie\, \left[k\cdot \bar A,
{\cal F}^{(0)}\right] = \frac{e}{2(2\pi)^{3}} \frac{\partial}{\partial
k_{\sigma}} k^{\rho}\left(\bar f_{\rho\sigma} (x)\star {\cal F}^{(0)} +
{\cal F}^{(0)}\star \bar f_{\rho\sigma} (x) - 8\delta (k^{2}) n_{B}
(|k^{0}|) \bar f_{\rho\sigma} (x+\theta\, k)\right)
\end{equation}
Since ${\cal F}^{(0)}$ is independent of $x$, its Moyal bracket with
$k\cdot \bar A (x)$ vanishes and using the form for ${\cal F}^{(0)}$ in
(\ref{f0final}), we determine
\begin{equation}
{\cal F}^{(1)} (x,k) = \frac{4e}{(2\pi)^{3}}\,\frac{1}{k\cdot
\partial}\frac{\partial}{\partial k_{\sigma}} \left(\delta (k^{2})
n_{B} (|k^{0}|) k^{\rho} (\bar f_{\rho\sigma} (x) - \bar f_{\rho\sigma}
(x+\theta\, k)\right)\label{f1final}
\end{equation}
\ewt

Before going onto calculate ${\cal F}^{(2)}$, let us discuss some of
the features of the results in (\ref{f0final}) and (\ref{f1final}). We
note that ${\cal F}^{(0)} (x,k)$ is manifestly covariant under a gauge
transformation, (\ref{gaugetfn1}). However, even though we start from
a gauge covariant equation, (\ref{fequation}), we see signs of
violation of gauge covariance in the calculation of ${\cal F}^{(1)}
(x,k)$. This is manifest more clearly in the right hand side of
(\ref{f1'}). Namely, we note that
\begin{equation}
e^{ik\cdot x}\star \bar f_{\rho\sigma} (x)
\star e^{-i k\cdot x}
\end{equation}
does not transform under a gauge transformation, (\ref{gaugetfn1}), as
we would expect. In a non-commutative gauge theory, space-time
translations form a subgroup of the gauge group and, in particular,
because of relations like (\ref{identity}), gauge covariance appears
to be violated. We want to emphasize that gauge covariance is manifest
before taking the ensemble average. However, the naive mean field ensemble
average seems to be incompatible with gauge covariance. It is worth
pointing out that this is not a problem in the usual QCD where factors
such as $e^{\pm ik\cdot x}$ are ordinary functions. This, therefore,
is a very special feature of the non-commutative nature of the theory and
implies that, in such theories, the naive mean field ensemble average
must be modified.

For lack of a more fundamental understanding of the mean field average
method in such theories, we proceed as follows. We note from
(\ref{identity}) that the ordinary plane wave function in
non-commutative theories leads to a translation which does not commute
with gauge transformation. The simplest way to covariantize such an
expression would be to replace the coordinate in the exponent of the
plane wave by a covariant coordinate which would generate a covariant
translation. Such a covariant coordinate can, in fact, be uniquely
determined to the leading order, from a few general conditions
that we discuss in detail in appendix {\bf B}. For the present,
however, we simply note that we
can uniquely identify the covariant coordinate with
\begin{equation}
X^{\mu} = x^{\mu} + {e} \theta^{\mu\nu}
\tilde{A}_{\nu} (x)\label{covariantcoordinate}
\end{equation}
where we identify (we will discuss this point more in detail in
appendix {\bf B})
\begin{equation}
\tilde{A}_{\mu} (x) = \bar A_{\mu} (x) + 
\frac{1}{k\cdot \bar D} \bar F_{\mu\nu} (x) k^{\nu}\label{atilde}
\end{equation}
Thus, we see that the simplest way to covariantize the mean field
calculations in non-commutative theories is to replace
\begin{equation}
e^{\pm ik\cdot x}\rightarrow e_{\star}^{\pm i(k\cdot x +
{e} k \times \tilde{A} (x))}\label{covariantization}
\end{equation}
where we have used the standard notation of non-commutative theories,
\begin{equation}
A\times B = \theta^{\mu\nu} A_{\mu}B_{\nu}
\end{equation}
We note that such a covariantization vanishes in the usual Yang-Mills
theories simply because $\theta^{\mu\nu}=0$, but is crucial for the
covariantization of the results in a non-commutative theory.

The covariantization in (\ref{covariantization}) contributes only at
higher orders in ``$e$''. Therefore, ${\cal F}^{(1)} (x,k)$ is
unaffected by this. However, in calculating ${\cal F}^{(2)} (x,k)$, we
have to take into account contributions coming from this in order to
maintain covariance under a gauge transformation. At order
``$e^{2}$'', the leading order contributions from the second term on
the right  hand side of (\ref{fequation}) have the forms
\bwt
\begin{eqnarray}
&  & {2e} \frac{\partial}{\partial k_{\sigma}} k^{\rho}\int
\frac{d^{4}y}{(\norma)^{4}}\,e^{-i y\cdot
k}\frac{1}{k^{2}}\nonumber\\
 &  & \quad \times\left\{ \langle (\partial_{\mu}a_{\lambda} (x_{+})
- \partial_{\lambda} a_{\mu} (x_{+}))\star \bar f_{\rho\sigma} (x)\star
\partial^{\lambda}a^{\mu} (x_{-})\rangle_{\rm cov}\right.\nonumber\\
 &  & \qquad -ie\, \langle (\partial_{\mu} a_{\lambda}
(x_{+}) - \partial_{\lambda} a_{\mu} (x_{+}))\star \left[\bar A_{\rho} (x),
\bar A_{\sigma} (x)\right]_{\rm MB}\star \partial^{\lambda}a^{\mu}
(x_{-})\rangle\nonumber\\
 &  & \qquad -ie\, \langle \left(\left[\bar A_{\mu}, a_{\lambda}
(x_{+})\right]_{\rm MB} - \left[\bar A_{\lambda}, a_{\mu}
(x_{+})\right]_{\rm MB}\right)\star \bar f_{\rho\sigma} (x)\star
\partial^{\lambda} a^{\mu} (x_{-})\rangle\nonumber\\
 &  & \left. - ie\, \langle (\partial_{\mu} a_{\lambda}
(x_{+}) 
- \partial_{\lambda} a_{\mu} (x_{+}))\star \bar f_{\rho\sigma} (x)\star
\left[\bar A^{\lambda}, a^{\mu} (x_{-})\right]_{\rm
MB}\rangle\right\}\label{f2}
\end{eqnarray}
\ewt
Here, $\langle \cdots \rangle_{\rm cov}$ stands for the linear terms
in $\bar A$ coming from the covariantization discussed in
(\ref{covariantization}). The other terms are already of order
``$e^{2}$'' so that the covariantization does not contribute in such
terms at this order. Equation (\ref{f2}) can be evaluated in a
straightforward manner to give
\bwt
\begin{equation}
\frac{4ie^{2}}{(2\pi)^{3}} \frac{\partial}{\partial
k_{\sigma}} 
k^{\rho} \left(\delta (k^{2}) n_{B} (|k^{0}|)\left\{ 
\left[\frac{1}{k\cdot \partial} k\cdot 
(\bar A(x+\theta\, k)-\bar A(x)),\bar f_{\rho\sigma}(x+
\theta\, k)\right]_{\rm MB} +
\left[\bar A_{\rho}
(x+\theta\, k), \bar A_{\sigma}
(x+{\theta k})\right]_{\rm MB}\right\} + \cdots\right)
\end{equation}
where ``$\cdots$'' represent terms that are of sub-leading
order. Substituting this into (\ref{fequation}), we can now determine
to leading order,
\begin{eqnarray}
{\cal F}^{(2)} (x,k) & = & ie\, \frac{1}{k\cdot
\partial} \left[k\cdot \bar A(x), {\cal F}^{(1)} (x,k)\right]_{\rm
MB}\nonumber\\ 
 &  & \quad -\frac{4ie^{2}}{(2\pi)^{3} } \frac{1}{k\cdot
\partial} \frac{\partial}{\partial k_{\sigma}} k^{\rho}
\Bigl\{\delta
(k^{2})n_{B} (|k^{0}|)\Bigl(\left[\bar A_{\rho} (x), \bar A_{\sigma}
(x)\right]_{\rm MB} - \left[\bar A_{\rho} (x+\theta\, k),
\bar A_{\sigma} (x+\theta\, k)\right]_{\rm
MB}\Bigr.\Bigr.\nonumber\\ 
 & & \quad\Bigl.\Bigl. +
\left[\frac{1}{k\cdot \partial} k\cdot (\bar A(x)-\bar 
A(x+\theta\, k)),\bar f_{\rho\sigma}(x+
\theta\, k)\right]_{\rm MB}\Bigr)\Bigr\}\label{f2final}
\end{eqnarray}
\ewt

This is all we need to determine the leading amplitudes up to the three
point function in the hard thermal loop approximation. However, let us
first note some of the basic features of these results. We note from
(\ref{f1final}) and (\ref{f2final}) that these Wigner functions have
the dipole structure characteristic of non-commutative theories. As we
have alluded to in the introduction, the constituents of
non-commutative theories can be thought of as extended particles and
have a dipole structure in the charged sector. This is basically
reflected in these calculations of the Wigner functions.

Given ${\cal F}^{(i)} (x,k), i=0,1,2$, we can now construct the
current up to third order in the coupling constant from the definition
in (\ref{current}). In momentum space, they have the explicit forms
\bwt
\begin{subequations}\label{currents}
\begin{equation}\label{currentsA}
J_{\mu}^{(0)} (-p_{1}) = 0
\end{equation}
\begin{equation}\label{currentsB}
J_{\mu}^{(1)} (-p_{1}) =  - {8e^{2}} \int
dK\,n_{B}(k^{0})\,(1- \cos ({p_{1}\times k})) L_{\mu\sigma}
(p_{1},k) L^{\nu\sigma} (p_{1},k) \bar A_{\nu} (-p_{1})
\end{equation}
\begin{eqnarray}\label{currentsC}
J_{\mu}^{(2)} (-p_{1}) & = &  {16ie^{3}} \int
d^{4}p_{2}d^{4}p_{3} dK\,n_{B} (k^{0}) \delta (p_{1}+p_{2}+p_{3}) 
\sin\left(\frac{p_{1}\times p_{2}}{2}\right)\nonumber\\
 &  & \quad \times\left[(1- \cos ({p_{3}\times k}))
 \left(L_{\mu\sigma} (p_{1},k) \frac{k_{\nu}}{p_{1}\cdot k} +
L_{\nu\sigma} (p_{3},k) \frac{k_{\mu}}{p_{1}\cdot k}\right)\bar A^{\nu}
(p_{2})  L^{\lambda\sigma}
(p_{3},k) \bar A_{\lambda} (p_{3})\right.\nonumber\\
&  & \qquad + (1- \cos ({p_{1}\times k})) \frac{k\cdot \bar A
(p_{3})}{p_{1}\cdot k} L_{\mu\sigma}(p_1,k) L^{\nu\sigma}(p_1,k)
\bar A_{\nu} (p_{2})\nonumber\\
 &  & \qquad \left. + (\cos ({p_{1}\times k}) - \cos
({p_{3}\times k})) \frac{p_{3}\cdot k}{p_{1}\cdot k}
\frac{k\cdot \bar A (p_{2})}{p_{2}\cdot k} L_{\mu\sigma} (p_{1},k)
L^{\lambda\sigma} (p_{3},k) \bar A_{\lambda} (p_{3})\right]
\end{eqnarray}
\end{subequations}
\ewt
where we have defined
\begin{subequations}\label{definitions}
\begin{equation}
dK = \frac{d^{4}k}{(2\pi)^{3}}
\,\theta (k^{0})\,\delta (k^{2})
\end{equation}
\begin{equation}
L_{\mu\nu} (p,k) = \eta_{\mu\nu} - \frac{k_{\mu}p_{\nu}}{p\cdot k}
\end{equation}
\end{subequations}
It is interesting to note here that even though the Wigner functions,
${\cal F}$, have a dipole structure, the odd nature of the current
under charge conjugation has led to a manifest quadrupole structure
for the currents. This was already noted in our earlier paper and here
we see explicitly the reason for this.

The currents are functionals of the gauge fields and, therefore, by
successively differentiating the current with respect to gauge fields,
we can determine various amplitudes. For example, the two point
function follows from
\bwt
\begin{equation}
\Pi_{\mu\nu} (p) = \left.\frac{\delta J_{\mu}
(-p)}{\delta \bar A^{\nu} (p)}\right|_{\bar A=0} = \frac{\delta J_{\mu}^{(1)}
(-p)}{\delta \bar A^{\nu} (p)} 
 =  -\frac{8\,e^2}{(2\pi)^3}\int {\rm d}K n_B(k^0)\, 
\left(1-\cos({p\times k})\right)
\,G_{\mu\nu}(p;\,k)\label{gaugeselfenergy}
\end{equation}
where we have defined
\be
G_{\mu\nu}(p;\,k)= \eta_{\mu\nu} - \frac{k_\mu\, p_\nu + k_\nu\,
p_\mu}{(k\cdot p)} + \frac{p^2\, k_\mu\, k_\nu}{(k\cdot p)^2} =
L_{\mu\sigma} (p,k) L_{\nu}^{\,\sigma} (p,k)\label{G} 
\ee
Similarly, the leading order three point amplitude is obtained to be
\begin{eqnarray}
\Gamma_{\mu\nu\lambda}(p_1,p_2,p_3) & = & \left.
\frac{\delta^{2} J_{\mu}(-p_{1})}
{\delta \bar A^{\nu} (p_{2})\delta \bar A^{\lambda}(p_{3})}\right|_{\bar A=0} = 
\frac{\delta^{2} J_{\mu}^{(2)}(-p_{1})}
{\delta \bar A^{\nu} (p_{2})\delta \bar A^{\lambda} (p_{3})}\nnbb
  & = & 
\frac{16\,i\,e^3}{(2\pi)^3}\,\sin\left(\frac{p_1\times p_2}{2}\right)
\int{\rm d}K\, n_B(k^0)\frac{1}{k\cdot p_1}
\left\{\left[1-\cos({p_1\times k})\right]
\,k_\lambda\,G_{\mu\nu}(p_1;\,k)
{{~}\atop{~}}\right.\nnbb
& & \qquad + \left[1-\cos({p_3\times k})\right]
\left[k_\mu\,G_{\nu\lambda}(p_3;\, k) + 
k_\nu\,G_{\mu\sigma}(p_1;\, k)\,G^\sigma_\lambda(p_3;\, k) \right] \nnbb
& & \quad + 
\left[\cos({p_1\times k})-\cos({p_3\times k})\right]
\left.\frac{k\cdot p_3}{k\cdot p_2}\,
k_\nu\,G_{\mu\sigma}(p_1;\, k)\,G^\sigma_\lambda(p_3;\, k)-
(p_2\leftrightarrow p_3; \nu\leftrightarrow \lambda)\right\}.
\end{eqnarray}
\ewt

These amplitudes agree completely with the leading order perturbative
results, arising from the gauge and the ghost loops, in the hard
thermal loop  approximation discussed in \cite{Brandt:2002qk}.
They are gauge independent and satisfy simple Ward
identities. For example, from the structure in (\ref{G}), one can
easily  verify the transversality of the photon self-energy
\be
p^\mu\, \Pi_{\mu\nu}(p) = 0,
\ee
Similarly, the identity relating the two and the three
point functions 
\bwt
\be
p_3^\lambda\,\Gamma_{\mu\nu\lambda}(p_1,p_2,p_3) = 2\,i
{e}\sin\left(\frac{p_1\times p_2}{2}\right)
\left[\Pi_{\mu\nu}(p_1)-\Pi_{\nu\mu}(p_2)\right].
\ee
can also be seen to hold.
\ewt

\section{Summary and discussions}

In this paper, we have studied the Wigner function approach to the
transport equation for photon in non-commutative QED. This is
complementary to our earlier work \cite{Brandt:2002qk}
in that here we have
tried to give a theoretical derivation for the transport
equation. While non-commutative QED has many similarities with the
conventional QCD, some differences arise in this method. In
particular, we have shown that because the gauge symmetry in
non-commutative QED is intermixed with space-time translations, the
naive mean field ensemble average breaks down and needs
modification. While we have proposed a covariantization with an
interest in the leading order hard thermal loop calculations, a
systematic study of this question remains to be carried out. We have
shown that the leading order amplitudes resulting from the current in
the  Wigner function transport equation approach in non-commutative
QED agree completely with the explicit perturbation calculations up to
the three point amplitudes.

We can also derive the force law for a neutral non-commutative
particle from our calculations. Let us recall that in the usual
description of the transport equation (using dynamical equations for
the particles), the current is defined as
\begin{equation}
J_{\mu} (x) = e\int dK\,k_{\mu} f(x,k)
\end{equation}
where $f(x,k)$ represents the distribution function. Comparing with
our definition of the current in (\ref{current}), we see that 
\begin{equation}\label{identification}
f(x,k)\sim  \frac{1}{2}({\cal F} (x,k) - {\cal F}(x,-k)).
\end{equation}
We may now relate the form of the transport equation (\ref{fequation})
with the structure based on the dynamical equations (\ref{force})
for particles in non-commutative QED. As shown in 
\cite{Brandt:2002qk}, to leading order, such a transport equation
would be of the form
\be\label{64}
k\cdot \bar D f(x,k) +
e\,\frac{\partial}{\partial k^\sigma}
\left(n_B(|k^0|)\, X^\sigma\right) =0.
\ee
Comparing (\ref{64}) with (\ref{fequation}), and using the
identification in (\ref{identification}), as well as the relations
(\ref{f1final}) and (\ref{f2final}), we find that
\begin{equation}
X^{\sigma} = {2} \left(1 - \cos k\times (i\tilde{D})\right)
\bar F^{\sigma\rho} k_{\rho}\label{force1}
\end{equation} 

This force equation for the charge neutral particle, agrees
completely with what we had obtained in our earlier paper. As we had
emphasized in that paper, we see that the nature of the force
on such a constituent is that of a quadrupole. However, this
theoretical derivation clarifies how dipole structures inherent in a
non-commutative theories give rise to a quadrupole effect. 
To understand this, we remark that up to corrections of order $e$, the
above result may be written in the form
\be
X^{\sigma} \simeq \left\{{k_\rho}
\left[\bar f^{\sigma\rho}(x) - 
      \bar f^{\sigma\rho}(x+\theta\, k)\right]\right\} -
\Bigl\{k\rightarrow -k\Bigr\}  .
\ee
We see that each of the above force terms indeed has a dipole
structure. However, the charge conjugation odd nature of the current
converts the dipoles into a quadrupole which effectively has the
structure of a pair of dipoles back to back.

\begin{acknowledgments}
We would like to thank Professor J.~C.~Taylor for helpful discussions.
This work was supported in part by US DOE Grant number DE-FG
02-91ER40685 and by CNPq and FAPESP, Brazil.
\end{acknowledgments}

\appendix

\section{Derivation of the transport equation}

In this appendix, we will give some details on the derivation of the
transport equation in (\ref{equation}). First, we note that, for a
link operator along a straight path, we can write,
\begin{eqnarray}
U (x, x_{\pm}) & = & {\rm P}\left(e_{\star}^{\frac{ie}{\hbar c}
\int_{0}^{1} dt\,\dot{\xi}(t)\cdot A (x+\xi(t))}\right) \nonumber \\
               & = & {\rm P}\left(e_{\star}^{\mp \frac{ie}{\hbar c}
\int_{0}^{1} dt\,\frac{y}{2}\cdot A (x \pm (1-t) \frac{y}{2})}\right)
\end{eqnarray}
where we have identified
\begin{equation}
\xi^{\mu} (t) = \pm (1-t) \frac{y^{\mu}}{2}
\end{equation}
It follows now from its definition that, under a general change of the
end points,  the link operator will change as
\bwt
\begin{equation}
\delta U (x, x_{\pm}) = \frac{ie}{\hbar c} \int_{0}^{1} dt\,U (x,
x+\xi(t))\star \delta
\left(\dot{\xi}(t)\cdot A(x+\xi(t))\right)\star U(x+\xi(t), x_{\pm})
\end{equation}
With some algebraic manipulations involving integration by parts, the
change can be rewritten in the form
\begin{eqnarray}
\delta U (x, x_{\pm}) & = & \frac{ie}{\hbar c} \delta x^{\mu}
\left(A_{\mu} (x)\star U
(x, x_{\pm}) - U (x, x_{\pm})\star A_{\mu} (x_{\pm})\right) \mp
\frac{ie}{2\hbar c} \delta y^{\mu} U (x, x_{\pm})\star A_{\mu}
(x_{\pm}) \nonumber\\
 & - &  \frac{ie}{\hbar c} \int_{0}^{1} dt\,\delta
(x+\xi)^{\mu} \left[\frac{dU (x, x+\xi)}{dt}\star A_{\mu} (x+\xi)\star
U(x+\xi, x_{\pm}) + U (x, x+\xi)\star A_{\mu} (x+\xi)\star \frac{dU
(x+\xi, x_{\pm})}{dt}\right.\nonumber\\
 &  & \qquad \left. - \dot{\xi}^{\nu} U (x, x+\xi)\star
(\partial_{\mu} A_{\nu} (x+\xi) - \partial_{\nu} A_{\mu} (x+\xi))\star
U (x+\xi, x_{\pm})\right]\label{derivation1}
\end{eqnarray}

Let us next recall that, by definition,
\begin{eqnarray}
U (x, x+\xi (t)) & = & {\rm P}\left(e_{\star}^{\frac{ie}{\hbar c}
\int_{t}^{1} dt'\,\dot{\xi}(t')\cdot A (x + \xi
(t'))}\right)\nonumber\\
U (x+\xi (t), x_{\pm}) & = & {\rm P}\left(e_{\star}^{\frac{ie}{\hbar
c} \int_{0}^{t} dt'\,\dot{\xi}(t')\cdot A (x+\xi(t'))}\right)
\end{eqnarray}
It follows from this that
\begin{eqnarray}
\frac{d U (x, x+\xi(t))}{dt} & = & - \frac{ie}{\hbar c} U (x,
x+\xi(t))\star \dot{\xi}(t)\cdot A (x+\xi(t))\nonumber\\
\frac{d U (x+\xi(t), x_{\pm})}{dt} & = & \frac{ie}{\hbar c}
\dot{\xi}(t)\cdot A (x+\xi(t))\star U(x+\xi(t), x_{\pm})
\end{eqnarray}
Using these relations, the variation in (\ref{derivation1}) can be
written in the simpler form
\begin{eqnarray}
\delta U (x, x_{\pm}) & = & \frac{ie}{\hbar c} \delta x^{\mu}
\left(A_{\mu} (x)\star U
(x, x_{\pm}) - U (x, x_{\pm})\star A_{\mu} (x_{\pm})\right) \mp
\frac{ie}{2\hbar c} \delta y^{\mu} U (x, x_{\pm})\star A_{\mu}
(x_{\pm})\nonumber\\
 &  & \quad - \frac{ie}{\hbar c} \int_{0}^{1} dt\,\dot{\xi}^{\mu}
(t)\delta(x+\xi)^{\nu} U (x, x+\xi(t))\star F_{\mu\nu} (x+\xi(t))\star
U(x+\xi(t), x_{\pm})\label{derivationfinal}
\end{eqnarray}
It now follows from this that, for $\delta y^{\mu} = 0$ and an
infinitesimal translation of $x^{\mu}$, we have
\begin{equation}
D^{(x)}_{\mu} U (x, x_{\pm}) = \mp \frac{i}{2}
y^{\nu}\left(\int_{0}^{1}dt\left(e_{\star}^{\pm \frac{ty}{2}\cdot D}
F_{\mu\nu} (x)\right)\right)\star U (x, x_{\pm})
\end{equation}
where the covariant derivative of the link operator is defined in
(\ref{relations}) and we have used the group properties of the link
operators in writing the expression in the final form. The second
relation in (\ref{relations}) can also be derived in an exactly
similar manner. Given these relations, the transport equation
(\ref{equation}) follows from the observation that
\begin{eqnarray}
D^{(x)}_{\rho} G^{(\pm)}_{\mu\nu} (x) & = & \left(D_{\rho}^{(x)} U (x,
x_{\pm})\right)\star F_{\mu\nu} (x_{\pm})\star U (x_{\pm}, x) + U(x,
x_{\pm})\star (D_{\rho}^{(x)}F_{\mu\nu} (x_{\pm}))\star U(x_{\pm},
x)\nonumber\\ 
&  & \qquad + U (x, x_{\pm})\star F_{\mu\nu} (x_{\pm})\star
\left(D_{\rho}^{(x)}U (x_{\pm}, x)\right)
\end{eqnarray}

The simplification of the second term in (\ref{equation}) presented in
(\ref{higherorder}) can be obtained as follows. First, we note that
the factor $k^{\rho}$ can be taken inside the integral as
\begin{equation}
k^{\rho}\rightarrow i\hbar \frac{\partial}{\partial y_{\rho}}
\end{equation}
Furthermore, the $y$-derivatives can be converted to covariant
derivatives and, from (\ref{derivationfinal}), we see that we can
write 
\begin{eqnarray}
D_{\mu}^{(y)} U (x, x_{\pm}) & = & \partial_{\mu}^{(y)} U (x, x_{\pm})
\pm \frac{ie}{2\hbar c} U (x, x_{\pm})\star A_{\mu}
(x_{\pm})
 =  - \frac{ie}{4\hbar c} y^{\nu}\left(\int_{0}^{1}
dt\,t\left(e_{\star}^{\pm \frac{ty}{2}\cdot D} F_{\mu\nu}
(x)\right)\right)\star U (x, x_{\pm})\nonumber\\
D_{\mu}^{(y)} U (x_{\pm}, x) & = & \partial_{\mu}^{(y)} U (x_{\pm}, x)
\mp \frac{ie}{2\hbar c} A_{\mu} (x_{\pm})\star U (x_{\pm},
x)
 =  \frac{ie}{4\hbar c} y^{\nu} U (x_{\pm}, x)\star
\left(\int_{0}^{1} dt\,t\left(e_{\star}^{\pm\frac{ty}{2}\cdot D}
F_{\mu\nu} (x)\right)\right)
\end{eqnarray}
\ewt
Using these as well as the identity (following from the Bianchi
identity as well as the equations of motion),
\begin{equation}
D^{2} F_{\mu\nu} = 2\,\frac{ie}{\hbar c} \left[F_{\mu\lambda},
F^{\lambda}_{\,\nu}\right]_{\rm MB}
\end{equation}
one obtains the expression given in Eq. (\ref{higherorder}).

\section{Some useful relations}

In this appendix, we will derive some other useful relations. First,
let us note from (\ref{xmoyal}) that 
\begin{eqnarray}
\left[x^{\mu}, f (x)\right]_{\rm MB} & = & i\theta^{\mu\nu}
(\partial_{\nu}f)\nonumber\\
{\rm or,}\quad \left[-i\theta^{-1}_{\mu\nu}x^{\nu}, f (x)\right]_{\rm
MB} & = & (\partial_{\mu}f)\label{derivative}
\end{eqnarray}
where we are assuming that $\theta^{\mu\nu}$ is invertible. Using
this, we see that we can write a covariant derivative (in the adjoint
representation)  with a gauge connection $\tilde{A}_{\mu}$ as
\begin{eqnarray}
(\tilde{D}_{\mu}f) & = & (\partial_{\mu}f) - \frac{ie}{\hbar c}
\left[\tilde{A}_{\mu}, f\right]_{\rm MB}\nonumber\\
& = &
\left[-i\left(\theta^{-1}_{\mu\nu}x^{\nu} + \frac{e}{\hbar c}
\tilde{A}_{\mu}\right), f\right]_{\rm MB}\nonumber\\
& = & \left[-i\theta^{-1}_{\mu\nu}\left(x^{\nu} + \frac{e}{\hbar c}
\theta^{\nu\lambda} \tilde{A}_{\lambda}\right), f\right]_{\rm
MB}\label{covariantderivative}
\end{eqnarray}
Here,
\begin{equation}
X^{\mu} = x^{\mu} + \frac{e}{\hbar c} \theta^{\mu\nu} \tilde{A}_{\nu}
(x)
\end{equation}
is known as the covariant coordinate, since its Moyal bracket leads to
the covariant derivative. Using relation (\ref{derivative})
repeatedly, it  follows that
\begin{equation}
e^{ik\cdot x}\star f (x)\star e^{-ik\cdot x} = \left(e^{-k\times
\partial}f\right)
\end{equation}
Similarly, using relation (\ref{covariantderivative}) repeatedly, it
follows that
\begin{eqnarray}
e_{\star}^{i(k\cdot x + \frac{e}{\hbar c} k\times \tilde{A})}\star
f\star e_{\star}^{-i(k\cdot x + \frac{e}{\hbar c} k\times \tilde{A})} 
\nonumber \\
 =  \left(e_{\star}^{-k\times \tilde{D}} f(x)\right) 
 =  \left(e_{\star}^{\tilde{k}\cdot \tilde{D}} f
(x)\right)\label{covarianttranslation}
\end{eqnarray}
where we have used the conventional definition (in non-commutative
field theories)
\begin{equation}
\tilde{k}^{\mu} = \theta^{\mu\nu} k_{\nu}
\end{equation}

Let us next present the arguments that determine the form of
$\tilde{A}_{\mu}$ uniquely. First, it must transform like a gauge
field and since the component of $\tilde{A}_{\mu}$ parallel to
$k^{\mu}$ does not enter into (\ref{covarianttranslation}), without
loss of generality, we can identify
\begin{equation}
k\cdot \tilde{A} = k\cdot \bar A\label{condn1}
\end{equation}
Thus, if we write
\begin{equation}
\tilde{A}_{\mu} = \bar A_{\mu} + Y_{\mu}
\end{equation}
it follows from (\ref{condn1}) that
\begin{equation}
k\cdot Y = 0\label{condn2}
\end{equation}
Since both $\bar A_{\mu}$ and $\tilde{A}_{\mu}$ transform like gauge
connections, it follows that $Y_{\mu}$ must transform covariantly in
the adjoint representation under a gauge transformation. If we now
assume that $Y_{\mu}$ has no explicit $\theta$ dependence except for
the ones coming from the star products (in order to be compatible with
charge conjugation), the form of $Y_{\mu}$ is determined to be
\begin{equation}
Y_{\mu} = H(k, \bar D, \bar F) \bar F_{\mu\nu} k^{\nu}\label{condn3}
\end{equation}
where $H$ is a gauge covariant function depending on $k^{\mu},
\bar D_{\mu}$ and $\bar F_{\mu\nu}$. 

Since $k^{2}=0$, on dimensional grounds, we can write
\begin{equation}
H (k, \bar D, \bar F) = h(k, \bar D, \bar F) 
\frac{1}{k\cdot \bar D} + g(k, \bar D, \bar F)
\frac{1}{\bar D^{2}}\label{condn4}
\end{equation}
where $h, g$ are dimensionless functions. We next make an important
observation following from the structure of thermal theories in the
hard thermal loop expansion. It is well known that in such a case only
singularities of the type $\frac{1}{(k\cdot \partial)^{n}}$
arise. Thus, if we are interested in making contact with thermal
field theories in the hard thermal loop approximation, the second
structure in (\ref{condn4}) cannot be present. Furthermore, if we
write
\begin{equation}
h (k, \bar D, \bar F) = \sum_{n=0} \frac{1}{(k\cdot \bar D)^{n}} C_{n}(k, \bar D, \bar F)
\end{equation}
with $C_{0}={\rm constant}$, it is clear that the higher order terms
in the series will be more and more sub-leading by power counting. If
we are only interested in the leading order terms, we can restrict
ourselves to the $n=0$ term in which case we can write
\begin{equation}
Y_{\mu} = C_{0} \frac{1}{k\cdot \bar D} \bar F_{\mu\nu}k^{\nu}\label{condn5}
\end{equation}
Finally, the constant $C_{0}$ is determined by requiring that the
current following from using such a $Y_{\mu}$ must be integrable (so
that it can be interpreted as coming from an effective action),
\begin{equation}
\frac{\delta J_{\mu}}{\delta \bar A^{\nu}} = \frac{\delta^{2}\Gamma}{\delta
\bar A^{\mu}\delta \bar A^{\nu}} = \frac{\delta J_{\nu}}{\delta \bar A^{\mu}}
\end{equation}
This uniquely determines the constant $C_{0} = 1$. It is interesting
that $C_{0}=0$ is ruled out by the integrability of the current. Thus,
we determine uniquely that
\begin{equation}
\tilde{A}_{\mu} = \bar A_{\mu} + \frac{1}{k\cdot \bar D} \bar F_{\mu\nu} k^{\nu}
\end{equation} 
which is the result used in (\ref{atilde}).

\bwt
Finally, let us indicate how one obtains terms linear in $\bar A_{\mu}$
from the exponents involving covariant coordinates in
(\ref{f2}). Using (\ref{covarianttranslation}) and setting $c=\hbar=1$, 
we note that to linear order in the gauge field, we can write
\begin{eqnarray}
e_{\star}^{i (k\cdot x + {e}
k\times \tilde{A})} &\star & \left. \bar f_{\rho\sigma} (x)\star
e_{\star}^{-i(k\cdot x + {e} k\times
\tilde{A})}\right|_{\rm lin} =
\left.\left(e_{\star}^{\tilde{k}\cdot \tilde{D}}
\bar f_{\rho\sigma} (x)\right)\right|_{\rm lin}\nonumber\\
 & = & - {ie}\sum_{n=1}^{\infty} \frac{1}{
n!}\left[\left\{(\tilde{k}\cdot\partial)^{n-1} (\tilde{k}\cdot
\tilde{A}) + (\tilde{k}\cdot\partial)^{n-2}(\tilde{k}\cdot
\tilde{A})(\tilde{k}\cdot\partial) + \cdots +
(\tilde{k}\cdot\tilde{A})(\tilde{k}\cdot\partial)^{n-1}\right\},\bar f_{\rho\sigma}\right]_{\rm
MB}\label{expansion}
\end{eqnarray}
This expression is best evaluated in momentum space, where the right
hand side takes the form
\begin{eqnarray}
& = & -{ie} \int d^{4}p_{2} d^{4}p_{3} \delta
(p_{1}+p_{2}+p_{3}) 
\left((-2i \sin \left(\frac{p_{2}\times p_{3}}{2}\right)\right))
(\tilde{k}\cdot \tilde{A} (p_{2})) \bar f_{\rho\sigma} (k_{3})\nonumber\\
 &  &\quad \times \sum_{n=1}^{\infty} \frac{1}{n!}
\left\{(i\tilde{k}\cdot p_{1})^{n-1} + (i\tilde{k}\cdot p_{1})^{n-2}
(-i\tilde{k}\cdot p_{3}) + \cdots + (-i\tilde{k}\cdot
p_{3})^{n-1}\right\}\nonumber\\
 & = & - {2e} \int d^{4}p_{2} d^{4}p_{3} \delta
(p_{1}+p_{2}+p_{3}) 
\sin\left(\frac{p_{1}\times p_{2}}{2}\right)
(\tilde{k}\cdot
\tilde{A} (p_{2})) \bar f_{\rho\sigma} (p_{3})\nonumber\\
 &  &\quad \times \sum_{n=1}^{\infty}
\frac{1}{n!}\,\frac{(i\tilde{k}\cdot p_{1})^{n} - (-i\tilde{k}\cdot
p_{3})^{n}}{(i\tilde{k}\cdot p_{1}) - 
(-i\tilde{k}\cdot p_{3})}\nonumber\\
 & = & - {2ie} \int d^{4}p_{2} d^{4}p_{3} \delta
(p_{1}+p_{2}+p_{3}) 
\sin\left(\frac{p_{1}\times p_{2}}{2}\right)
\frac{\tilde{k}\cdot
\tilde{A} (p_{2})}{\tilde{k}\cdot p_{2}} \bar f_{\rho\sigma} (p_{3})
\left(e^{i \tilde{k}\cdot p_{1}} - e^{-i
\tilde{k}\cdot p_{3}}\right)\label{expansionfinal}
\end{eqnarray}
\ewt
where
\begin{equation}
\bar f_{\rho\sigma} (p_{3}) = -i\left({p_3}_\rho \bar A_{\sigma} (p_{3}) -
{p_3}_\sigma \bar A_{\rho} (p_{3})\right)
\end{equation}
When charge conjugation is implemented in the current, the last
parenthesis in (\ref{expansionfinal}) becomes a difference of two
cosines, as is manifested in the last term of $J_{\mu}^{(2)}$
in (\ref{currents}).


\begin{thebibliography}{10}

\bibitem{kapusta:book89lebellac:book96das:book97}
J.~I. Kapusta, {\em Finite Temperature Field Theory} (Cambridge
University
  Press, Cambridge, England, 1989);\\
%
M.~L. Bellac, {\em Thermal Field Theory} (Cambridge University
Press,
  Cambridge, England, 1996); \\
%
A. Das, {\em Finite Temperature Field Theory} (World Scientific,
NY, 1997).
\bibitem{Braaten:1990it}
E. Braaten and R.~D. Pisarski, 
Nucl. Phys. {\bf B337,} 569 (1990);
%
Nucl. Phys. {\bf B339,} 310 (1990).
\bibitem{Kobes:1990dc}
R.~Kobes, G.~Kunstatter and A.~Rebhan,
Nucl.\ Phys.\ B {\bf 355}, 1 (1991).
\bibitem{frenkel:1991ts}
J. Frenkel and J.~C. Taylor, 
Nucl. Phys. {\bf B334,} 199 (1990); Nucl. Phys. {\bf B374,} 156 (1992).
\bibitem{Heinz:1985yq}
U.~W. Heinz, 
Ann. Phys. {\bf 161,} 48 (1985);
%
Ann. Phys. {\bf 168,} 148 (1986).
\bibitem{Jackiw:1993zr}
R.~Jackiw and V.~P.~Nair,
Phys.\ Rev.\ D {\bf 48}, 4991 (1993).
\bibitem{Nair:1993rx}
V.~P.~Nair,
Phys.\ Rev.\ D {\bf 48}, 3432 (1993);
Phys.\ Rev.\ D {\bf 50}, 4201 (1994).
\bibitem{Blaizot:1994be}
J.~P. Blaizot and E. Iancu, 
Nucl. Phys. {\bf B417,} 608 (1994);
%
Nucl. Phys. {\bf B421,} 565 (1994).
\bibitem{kelly:1994ig}
P.~F. Kelly, Q. Liu, C. Lucchesi, and C. Manuel, 
Phys. Rev. Lett. {\bf 72,} 3461 (1994);
Phys. Rev. {\bf D50,} 4209 (1994).
\bibitem{Litim:sh}
D.~F.~Litim and C.~Manuel,
Phys.\ Rept.\  {\bf 364}, 451 (2002).
%
\bibitem{Elze:1989un}
H.~T.~Elze and U.~W.~Heinz,
Phys.\ Rept.\  {\bf 183}, 81 (1989).
%
\\
H.~T.~Elze, M.~Gyulassy and D.~Vasak,
Phys.\ Lett.\ B {\bf 177} (1986) 402;
Nucl.\ Phys.\ B {\bf 276} (1986) 706.
\bibitem{Seiberg:1999vs}
N. Seiberg and E. Witten,
JHEP {\bf 09,} 032 (1999).
\bibitem{Fischler:2000fv}
W. Fischler, J. Gomis, E. Gorbatov, A. Kashani-Poor, S. Paban, and P. Pouliot,
JHEP {\bf 05,} 024 (2000);
W. Fischler, E. Gorbatov, A. Kashani-Poor, R. McNees, S. Paban, P. Pouliot, 
JHEP {\bf 06,} 032 (2000).
\bibitem{Arcioni:1999hw}
G. Arcioni and M.~A. Vazquez-Mozo, 
JHEP {\bf 01,} 028 (2000).
\bibitem{Landsteiner:2000bw}
K. Landsteiner, E. Lopez, and M.~H.~G. Tytgat, 
JHEP {\bf 09,} 027 (2000);
%
JHEP {\bf 06,} 055 (2001).
\bibitem{Szabo:2001kg}
R.~J.~Szabo,
Phys.\ Rept.\  {\bf 378}, 207 (2003).
\bibitem{Douglas:2001ba}
M.~R. Douglas and N.~A. Nekrasov, 
Rev. Mod. Phys. {\bf 73,} 977 (2002).
\bibitem{Chu:2001fe}
A.~Armoni, R.~Minasian and S.~Theisen,
Phys.\ Lett.\ B {\bf 513}, 406 (2001);
C.-S. Chu, V.~V. Khoze, and G. Travaglini, 
Phys. Lett. {\bf B513,} 200 (2001);
%
C.-S. Chu, V.~V. Khoze, and G. Travaglini,
hep-th/0112139 (2001).
\bibitem{VanRaamsdonk:2001jd}
M. Van~Raamsdonk,
JHEP {\bf 11,} 006 (2001).
\bibitem{Bonora:2000ga}
A.~Armoni,
Nucl.\ Phys.\ B {\bf 593}, 229 (2001);
L. Bonora and M. Salizzoni,
Phys. Lett. {\bf B504,} 80 (2001).
\bibitem{Brandt:2002aa}
F.~T.~Brandt, J.~Frenkel and D.~G.~C.McKeon,
Phys.\ Rev.\ D {\bf 65}, 125029 (2002).
\bibitem{Brandt:2002rw}
F.~T.~Brandt, A.~Das, J.~Frenkel, D.~G.~McKeon and J.~C.~Taylor,
Phys.\ Rev.\ D {\bf 66}, 045011 (2002).
\bibitem{Minwalla:1999px}
S. Minwalla, M. Van~Raamsdonk, and N. Seiberg, 
JHEP {\bf 02,} 020 (2000).
\bibitem{Sheikh-Jabbari:1999vm}
M.~M. Sheikh-Jabbari, 
Phys. Lett. {\bf B455,} 129 (1999).
\bibitem{Bigatti:1999iz}
D. Bigatti and L. Susskind,
Phys. Rev. {\bf D62,} 066004 (2000).
\bibitem{Brandt:2002qk}
F.~T.~Brandt, A.~Das and J.~Frenkel,
Phys.\ Rev.\ D {\bf 66}, 105012 (2002).
\bibitem{Sheikh-Jabbari:2000vi}
M.~M.~Sheikh-Jabbari,
Phys.\ Rev.\ Lett.\  {\bf 84}, 5265 (2000).
\end{thebibliography}
\end{document}